# Time-Stepped Cyber-Physical Simulation of DoS, DoD, and FDI Attacks on the IEEE 14-Bus System

Manuella Christelle Tossa*, Fernando Madrigal*, Ryan Blosser*, Asma Jodeiri Akbarfam*
* Washington State University Tri-Cities, USA
{manuella.tossa, fernando.madrigal, ryan.blosser, asma.akbarfam}@wsu.edu

*Abstract*—Reliable grid operation depends on accurate and timely telemetry, making modern power systems vulnerable to communication-layer cyberattacks. This paper evaluates how Denial-of-Service (DoS), Denial-of-Data (DoD), and False Data Injection (FDI) attacks disrupt the IEEE 14-bus system using a MATLAB-only, time-stepped simulation framework built on MATPOWER. The framework emulates a 24-hour operating cycle with sinusoidal load variation, introduces attack-specific manipulation of load and voltage data, and performs full AC power-flow solves with reactive-limit enforcement (PV–PQ switching). At each timestep, the system logs true and measured voltages, generator P/Q output, system losses, and voltage-limit violations to capture transient cyber–physical effects. Results show that DoD causes the largest physical distortions and reactive-power stress, DoS masks natural variability and degrades situational awareness, and FDI creates significant discrepancies between true and perceived voltages. The study provides a compact, reproducible benchmark for analyzing cyber-induced instability and informing future defense strategies.

## I. INTRODUCTION

Modern power systems increasingly rely on digital communication networks and automated control, making them vulnerable to cyberattacks that compromise measurement and telemetry channels. Grid monitoring depends on accurate, real-time data from sensors, meters, and remote terminal units (RTUs). When these data streams are withheld, delayed, or corrupted, operators and control algorithms may misinterpret system conditions, potentially causing voltage violations, improper dispatch decisions, or cascading outages [1]–[3].

Communication-layer attacks such as Denial-of-Service (DoS), Denial-of-Data (DoD), and False Data Injection (FDI) directly target the availability and integrity of power-system telemetry [4]–[6]. DoS attacks block or delay measurement updates, forcing reliance on stale information; DoD attacks falsify reported load values, distorting local power balance; and FDI attacks bias voltage measurements, misleading state estimation and masking or fabricating anomalies. These threats align with adversarial models described in standards such as NIST SP800-82 and IEC 62351 [1], [7], underscoring their real-world relevance.

Despite extensive research on cyberattacks against power grids [7], [8], many studies rely on static steady-state analysis or simplified attack models. There remains a need for time-resolved cyber–physical simulations that capture how telemetry manipulation propagates through AC power-flow dynamics, reactive-power limits, and operator perception over time.

To address this gap, we develop a MATLAB-based, time-stepped cyber–physical simulation framework built on MATPOWER that separates true physical system states from corrupted telemetry. The framework emulates a 24-hour operating cycle with realistic load variation, injects DoS, DoD, and FDI attacks at the communication layer, and solves full AC power flow with reactive-limit enforcement and PV–PQ switching. Using this framework, results show that DoD attacks produce the largest physical disturbances and reactive-power stress, DoS suppresses natural variability and degrades situational awareness, and FDI primarily distorts perceived voltages while leaving the physical system state largely unchanged.

The primary contributions of this paper are as follows:
- A discrete-time cyber–physical simulation framework that integrates load variation, communication-layer attack models, and AC power-flow computation with PV–PQ switching.
- A systematic evaluation of DoS, DoD, and FDI attacks on voltages, generator behavior, reactive support, and overall grid stability in the IEEE 14-bus system.
- Quantitative and qualitative analysis using time-series voltages, heatmaps, RMS deviation, violation counts, and reactive-power trajectories to characterize attack-induced instability.
- A reproducible benchmark that supports future extensions involving realistic communication effects, detection mechanisms, and defensive controls.

This study demonstrates that even short-duration cyberattacks can create significant discrepancies between true and perceived grid conditions, highlighting the need for robust monitoring and cyber-resilient control strategies in modern power networks.

## II. RELATED WORK

Research on CYBSEC impacts to power-system operations commonly relies on small benchmark networks to enable controlled experimentation. Among these, the IEEE 14-bus in Figure 1system is one of the most frequently used T&D test cases due to its simple topology, mixed GEN/LD configuration, and suitability for studying measurement- and load-level perturbations [9]. Prior simulation studies have used this network to analyze both steady-state vulnerabilities and cyber-induced operating deviations. Cyberattacks on power-system telemetry are typically categorized into DoS, DoD, and FDI classes. NIST SP 800-82 describes how DoS attacks impair AVAIL by delaying or blocking CTRL msgs, while DoD corruption alters MEAS values such as bus LDs [1]. FDI attacks, originating from the seminal work of Liu *et al.*, intentionally bias TELEMETRY (e.g., V/P/Q) to evade BDD and mislead SE and control loops [5]. Recent studies have expanded FDI attack construction and detection, including black-box FDI methods [10], max–min optimization strategies [11], updated surveys on detection/localization [12], and new 2025 frameworks for state-space decomposition and ML-based detection under limited observability [13], [14]. These efforts reinforce the continued relevance of telemetry-manipulation threats but often focus on algorithmic or SE-level impacts rather than closed-loop PF behavior.

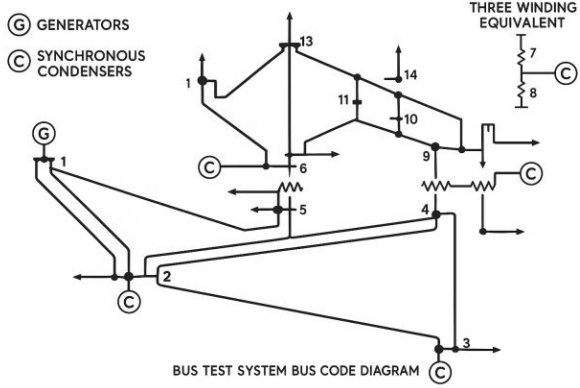

Fig. 1. IEEE 14-Bus System One-Line Diagram.

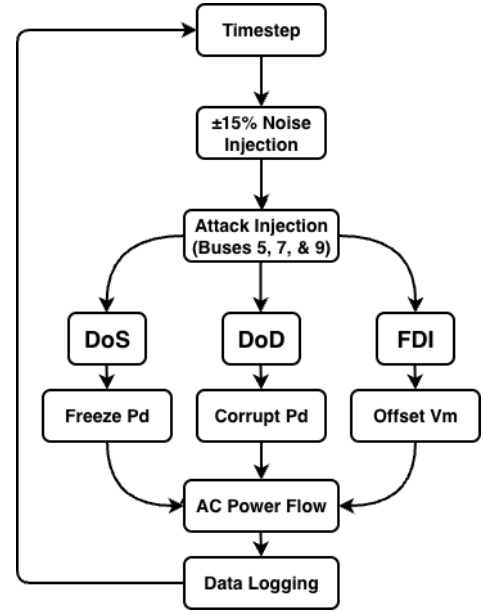

Fig. 2. Timestep workflow for cyberattack-enabled AC power flow simulation.

Analytical tools have played a central role in studying these disturbances. MATPOWER [15] is widely adopted for PF/OPF analysis and underpins many cyberattack-emulation studies, including DoS-induced MEAS loss and FDI-style load perturbations. However, existing work typically evaluates attacks using static PF snapshots, single-step perturbations, or simplified quasi-static evolutions. Even recent MATPOWER-based studies of DoS/FDI impacts do not incorporate full time-stepped interactions between evolving loads, telemetry manipulation, and reactive-power constraints.

Across these efforts, the dominant focus remains on steady-state or partially dynamic evaluations. To our knowledge, no existing framework injects DoS, DoD, and FDI attacks into a time-stepped PF environment while preserving system-level constraints such as PV–PQ switching, Q-limit enforcement, and generator reactive behavior. Our work addresses this gap by generating synchronized "true" and "measured" telemetry streams under cyberattack and quantifying the resulting V-behavior over time.

## III. SYSTEM DESIGN AND METHODOLOGY

This section outlines the simulation architecture, timestep workflow, cyberattack mechanisms, and threat model used to evaluate cyber–physical disturbances on the IEEE 14-bus system. The experiment is implemented in MATLAB using MATPOWER 5.0 and operates in discrete time, mimicking periodic SCADA telemetry updates and the possibility of compromised measurements.

### A. Overview of the Simulation Framework

The simulation framework is built around the IEEE 14-bus test system, a standard benchmark for voltage stability and power-flow studies. A discrete-time loop emulates a 24-hour operational cycle, during which the system experiences natural load variation, potential cyber disruptions, and AC power-flow computations with reactive-limit enforcement. This structure enables the study of transient cyber–physical interactions rather than static steady-state effects.

### B. Timestep Workflow

As illustrated in Fig. 2, each timestep consists of four major stages.

*1) Load Update:* A time-varying load multiplier is applied using a sinusoidal diurnal profile with ±15% ambient noise:

$$P_d(t) = P_{d0}\left(1 + 0.15\sin\frac{2\pi t}{24}\right),$$

capturing realistic fluctuations in system demand across the 24-hour cycle.

*2) Cyberattack Injection:* At each timestep, the simulation may introduce one of three communication-layer attack types:

- **Denial-of-Service (DoS):** The load vector $P_d$ is frozen at its previous value, emulating lost or delayed telemetry.
- **Denial-of-Data (DoD):** Load values at selected buses (5, 7, or 9) are scaled by a factor between 0.2 and 1.5, representing falsified field measurements.
- **False Data Injection (FDI):** A constant bias of +0.1 p.u. is added to measured voltages $V_m^{\text{meas}}$ after power-flow convergence, simulating corrupted telemetry received by the control center.

These attacks represent widely studied data-availability and data-integrity threats in power-system cybersecurity research.

*3) Power-Flow Solution:* A full AC power-flow calculation is performed using MATPOWER's Newton–Raphson solver (runpf). Reactive-limit enforcement is enabled via:

mpoption('enforce_q_lims', 1),

activating PV→PQ switching whenever a generator reaches its reactive-power boundary. This behavior approximates real-world voltage-regulation constraints under stressed or corrupted operating conditions.

*4) Data Logging:* For each timestep, the following variables are recorded:

- True and measured voltage magnitudes ($V_m^{\text{true}}$, $V_m^{\text{meas}}$)
- Generator real and reactive outputs ($P_g$, $Q_g$)
- Total load, total generation, and system losses
- Voltage-limit violations relative to IEEE 1547 (0.95–1.05 p.u.)
- Attack identifier bitmask (1 = DoS, 2 = DoD, 4 = FDI)

These logged values form the basis for later performance comparison and anomaly analysis.

*C. Threat Model*

The simulated adversary is assumed to have access to communication channels or Remote Terminal Units (RTUs), enabling manipulation of telemetry data. The threat model includes:

- **Motivation:** To disrupt system situational awareness or induce unstable operating conditions.
- **Capabilities:**
  - *Packet blocking (DoS):* disrupts data availability.
  - *Load falsification (DoD):* alters reported active-power demand $P_d$.
  - *Telemetry manipulation (FDI):* injects voltage biases into reported measurements.
- **Impact Path:** Corrupted or missing load and voltage measurements lead to inaccurate power-flow solutions, voltage imbalance, reactive-power stress, and potential PV→PQ switching or generator-limit violations.

In summary, the simulation advances in discrete timesteps by updating system loads according to a diurnal profile, injecting communication-layer attacks, solving the AC power flow with reactive-limit enforcement, and logging both true and measured telemetry. This structured workflow enables direct observation of how data loss or corruption propagates through power-flow behavior over time and affects both physical system states and operator perception.

TABLE I
SIMULATION PARAMETERS

| Parameter | Value |
| --- | --- |
| Timesteps | 144 (24-hour cycle) |
| Voltage Range | 0.95–1.05 p.u. (IEEE 1547) |
| Solver | Newton–Raphson (runpf) |
| PV–PQ Switching | Enabled (enforce_q_lims=1) |
| Load Profile | $1 + 0.15\sin(2\pi t/24)$ |
| Attack Bitmask | 1 = DoS, 2 = DoD, 4 = FDI |

IV. IMPLEMENTATION

The cyber–physical simulation was implemented in MATLAB R2023a using the MATPOWER 5.0 toolbox for steady-state AC power-flow analysis[1] Although MATPOWER 5.0 is used to maintain consistency with widely adopted IEEE benchmark studies, all simulations were executed in a modern MATLAB environment (R2023a), and the framework is not tied to a specific MATLAB release.

MATLAB served as the primary environment for algorithm development, timestep iteration, attack scheduling, and visualization, while MATPOWER provided standardized IEEE test cases, Newton–Raphson solvers (runpf), and structured access to bus, generator, and branch data structures required for realistic voltage-control behavior. MATPOWER 5.0 was selected to maintain consistency with commonly used benchmark studies of the IEEE 14-bus system. The proposed framework itself is not tied to a specific MATLAB release and relies only on standard AC power-flow functionality. Table I summarizes the key parameters governing the 24-hour simulation.

*A. IEEE 14-Bus System Configuration*

The study uses MATPOWER's IEEE 14-bus test system, which models a compact but realistic transmission network. The configuration includes:

- 14 buses (numbered 1–14),
- 5 generators located at buses 1, 2, 3, 6, and 8,
- Slack bus: bus 1,
- Loads distributed across buses 2–5 and 9–14,
- Transformers on branches 4–7, 4–9, and 5–6,
- Nominal base: 100 MVA, 60 Hz.

Baseline active-power demand ($P_d$) values were obtained from the MATPOWER case14 file. A sinusoidal load multiplier with ±15% noise was applied over 144 timesteps to emulate natural demand fluctuations consistent with the methodology described in Section III.

---

[1] Project code and simulation files: https://github.com/RyanWSU/CPTS455-simulation-of-DoS-DoD-and-FDI-attacks.

TABLE II
CYBERATTACK SCENARIOS

| Attack Type | Description | Window |
|---|---|---|
| DoS | Freeze $P_d$ (packet loss) | 20–50 |
| DoD | Scale $P_d$ at buses 5/7/9 by 1.5× | 100–130 |
| FDI | Add +0.1 p.u. bias to $V_m^{\text{meas}}$ | 60–90 |

### B. Cyberattack Implementation

The three attack types introduced in the timestep workflow (Section III) were integrated into the simulation loop as summarized in Table II. Each scenario evaluates the system's response to a different form of data corruption or loss.
Specifically:

- **Denial-of-Service (DoS, 20–50):** Communication loss is modeled by freezing the load vector $P_d$ at its last valid value, preventing telemetry updates.
- **Denial-of-Data (DoD, 100–130):** Selected buses (5, 7, or 9) have their active-power demand scaled by a factor of 1.5, representing falsified field measurements.
- **False Data Injection (FDI, 60–90):** A +0.1 p.u. bias is added to measured voltages $V_m^{\text{meas}}$ after power-flow convergence, distorting operator perception while leaving the physical voltages unchanged.

These attack windows correspond to realistic communication-layer disruptions described in NIST SP800-82 and IEC 62351.

### C. Power-Flow Simulation and Reactive-Limit Enforcement

AC power flow was solved at every timestep using the Newton–Raphson method. Reactive-power limits were enforced via

mpoption('enforce_q_lims', 1),

which activates PV→PQ switching whenever a generator reaches its reactive limit. This behavior reflects real-world voltage-control constraints during stressed conditions or corrupted input data.

For each timestep, the simulation recorded:
- true and measured voltages ($V_m^{\text{true}}$, $V_m^{\text{meas}}$),
- voltage angles ($\theta$),
- generator real and reactive outputs ($P_g$, $Q_g$),
- total load, total generation, and system losses,
- voltage-violation counts, and
- the attack identifier bitmask.

### D. Visualization and Analysis

Post-processing and visualization were performed in MATLAB. The following plots were generated to analyze system behavior under baseline and attacked conditions:
- bus-voltage time-series plots,
- voltage heatmaps (attacked, baseline, and delta),
- RMS voltage deviation traces for anomaly detection,
- attack-timeline overlays,

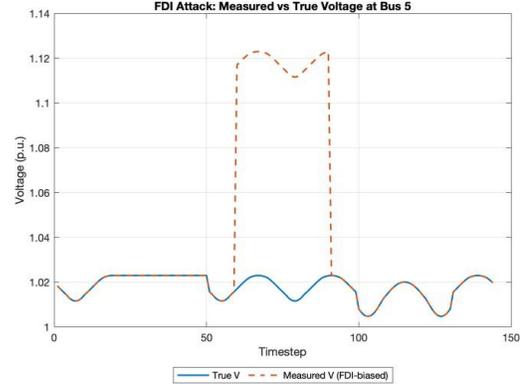

Fig. 3. FDI: measured vs. true voltages at bus 5.

- generator P/Q trajectories and PV→PQ switching indicators.

These visualizations show how communication-layer disruptions—whether via data loss, falsification, or voltage telemetry bias—propagate into measurable physical instability in the IEEE 14-bus system.

## V. RESULTS

The following subsections present the main results of the simulation, emphasizing how DoS, DoD, and FDI attacks affect bus voltages, generator behavior, and overall system stability.

### A. FDI: Measured vs. True Voltages at Bus 5

Figure 3 compares the measured and true voltages at bus 5 during the FDI attack window. As expected, the measured voltages exhibit a constant positive bias of +0.1 p.u. during timesteps 60–90, while the true voltages remain within the nominal band. This illustrates how FDI alters perceived system health without necessarily changing the underlying physical state.

### B. Selected Bus Voltages Under Attack

Figure 4 shows time-series voltage trajectories for representative buses (1, 5, 7, and 9). The top panel corresponds to the attacked run, and the bottom panel shows the baseline (no-attack) case. Shaded regions indicate the DoS (20–50), DoD (100–130), and FDI (60–90) windows, while dashed lines mark the nominal voltage range (0.95–1.05 p.u.).

During DoS, voltage curves flatten because load updates are frozen. DoD windows produce visible voltage jumps, particularly at bus 5, where falsified load inputs are applied. In contrast, the baseline run remains well-behaved and within limits, highlighting the disturbance introduced by cyberattacks.

### C. Voltage Heatmap Comparison

Figure 5 provides a heatmap comparison of bus voltages over time. The top panel shows attacked voltages, the middle panel shows the baseline, and the bottom panel displays the difference (attacked minus baseline). Rows correspond to bus indices and columns to timesteps; blue and red regions in the delta panel highlight undervoltages and overvoltages, respectively.

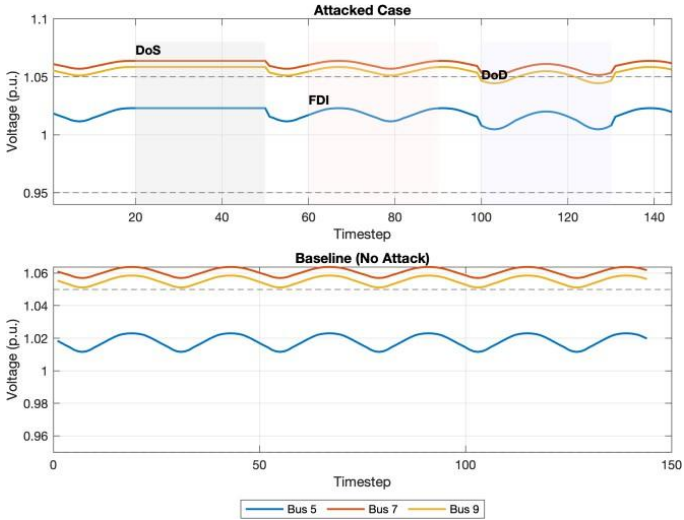

Fig. 4. Selected bus voltages (attacked vs. baseline).

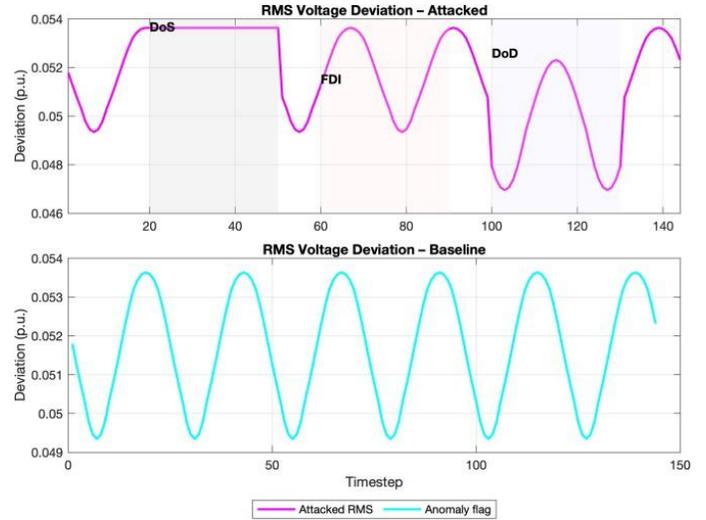

Fig. 6. RMS voltage deviation (attacked vs. baseline).

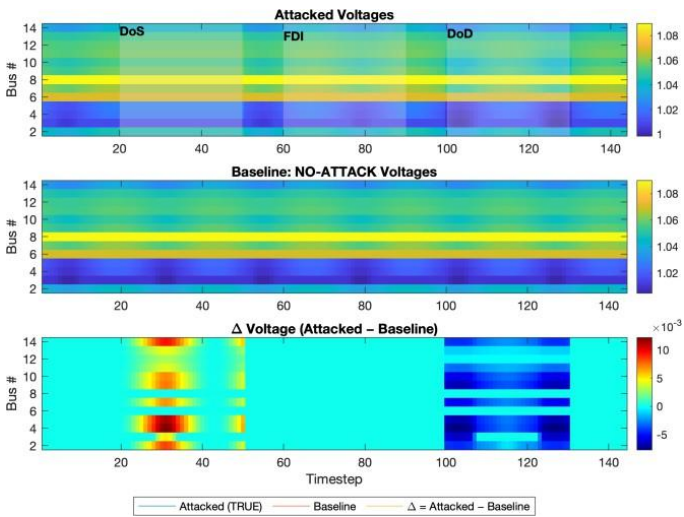

Fig. 5. Voltage heatmap comparison: attacked, baseline, and delta.

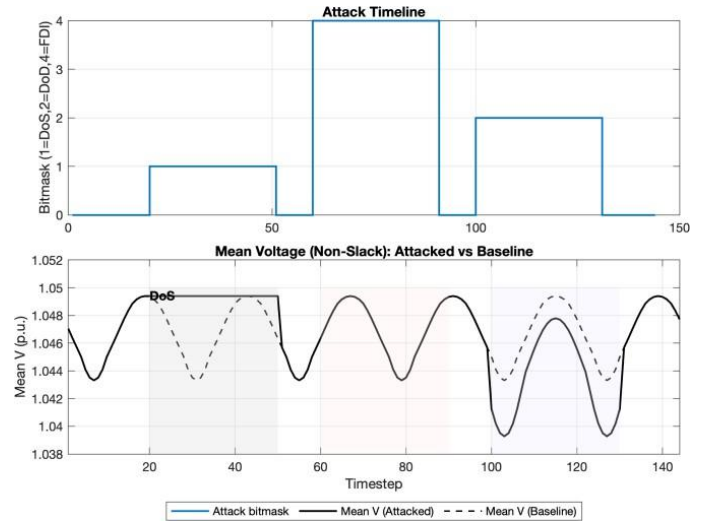

Fig. 7. Attack timeline and mean system voltage.

DoS periods exhibit nearly flat patterns, reflecting frozen telemetry. DoD intervals create localized red clusters near buses 5–9, where falsified loads distort the power balance. FDI effects appear briefly but system-wide, since corrupted telemetry influences perceived voltages across all buses. The delta map emphasizes how localized data corruption can propagate through the network.

### D. RMS Voltage Deviation

Figure 6 shows the RMS voltage deviation from 1 p.u. across all buses for the attacked and baseline runs. Markers denote detected anomalies. DoD windows generate the highest RMS spikes due to falsified load data, while DoS periods appear flatter, reflecting the effect of frozen measurements. FDI introduces smaller but noticeable increases in RMS deviation, capturing the measurement bias. RMS deviation thus serves as a compact indicator of system disturbance intensity.

### E. Attack Timeline and Mean Voltage

Figure 7 correlates the attack timeline with system-wide mean voltage. The top panel shows the attack bitmask (1 = DoS, 2 = DoD, 4 = FDI), and the bottom panel plots mean system voltage for the attacked and baseline runs. Mean voltage decreases slightly during DoS due to stagnant load data and increases during DoD when falsified low-load readings cause generators to raise voltage setpoints. The FDI window introduces a brief offset between true and perceived voltages. The alignment between attack markers and voltage deviations confirms the causal effect of data tampering.

### F. System Power Balance

Figure 8 compares total generation, total load, and system losses for attacked (solid) and baseline (dashed) runs. DoS periods show nearly constant curves because load and generation updates are frozen. During DoD, reduced (falsified) load readings create a mismatch in the apparent power balance, making generation appear excessive and causing fluctuations in system losses. This illustrates how integrity attacks distort the operator's perception of demand and can lead to inefficient or unstable control actions.

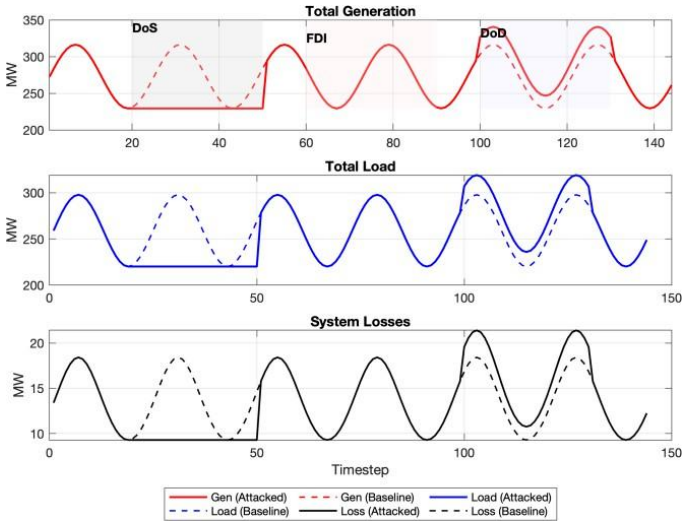

Fig. 8. System power balance (generation, load, and losses).

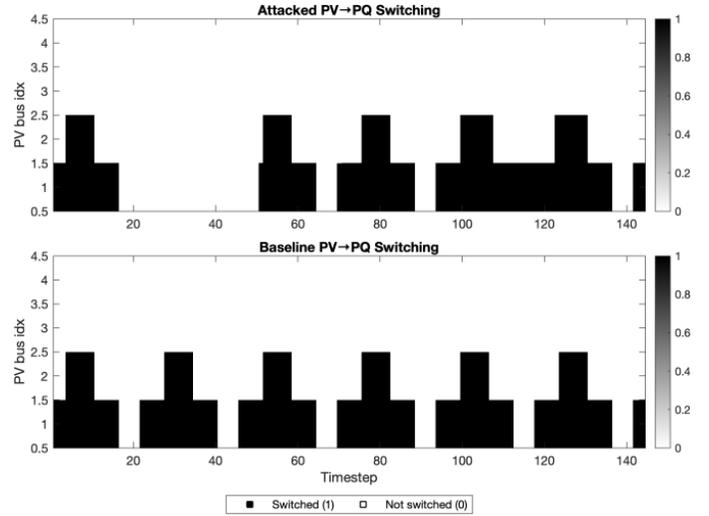

Fig. 10. PV→PQ switching events for attacked and baseline runs.

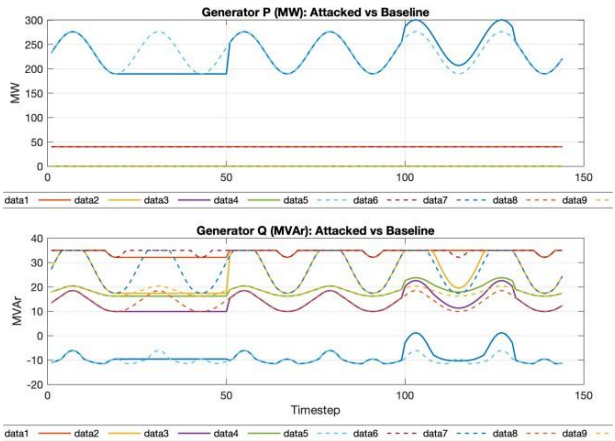

Fig. 9. Generator *P/Q* comparison (attacked vs. baseline).

### G. Generator P/Q Outputs and PV→PQ Switching

Figure 9 shows generator active ($P$) and reactive ($Q$) outputs for the attacked and baseline runs. Reactive-power traces reveal instances where generators hit their $Q$-limits, triggering PV→PQ switching events. These transitions are further summarized in Fig. 10, which plots binary indicators of PV→PQ status for attacked and baseline runs. Under normal conditions, switching is rare or absent; under attack, additional switching occurs, particularly during DoD windows when false load reductions force reactive overcompensation. Frequent switching indicates voltage-regulation stress and reactive-limit violations.

### H. Qualitative Analysis of Attack Impacts

The results demonstrate that even short-duration cyberattacks can induce measurable, system-wide physical impacts on the IEEE 14-bus system. Each attack type affects grid behavior through a different mechanism:

- **Denial-of-Service (DoS):** Primarily affects data availability by freezing load updates. This produces flattened voltage trajectories and delayed system response, as operators continue to rely on stale telemetry.
- **Denial-of-Data (DoD):** Directly affects data integrity by altering active-power demand values. DoD causes incorrect load balancing, voltage drift, and significant increases in generator reactive-power effort, leading to voltage violations and PV–PQ switching events.
- **False Data Injection (FDI):** Manipulates measurement accuracy by injecting false voltage telemetry. Although the physical grid may remain stable, the operator or estimator can be misled about grid health, potentially masking true violations or creating false alarms.

By enabling PV→PQ switching and enforcing generator reactive-power limits, the simulation captures nonlinear behaviors present in real transmission networks during stressed or compromised conditions, making the observed dynamics representative of realistic cyber–physical interactions.

### I. Quantitative Performance Metrics

Table III summarizes system-wide performance metrics for each attack scenario.

This comparison highlights three key observations:

1) DoD produces the largest physical deviations, reflected in higher voltage violations and increased losses.
2) DoS suppresses normal variability, reducing RMS deviation but causing operational blindness due to frozen telemetry.

TABLE III
SYSTEM PERFORMANCE UNDER ATTACK

| Metric | True | DoS | DoD | FDI |
|---|---|---|---|---|
| RMS Volt. Dev. (p.u.) | 0.0516 | 0.0536 | 0.0493 | 0.0519 |
| Max Volt. Dev. (p.u.) | 0.0536 | 0.0536 | 0.0523 | 0.0536 |
| Volt. Violations | 1014 | 248 | 182 | 224 |
| Avg. Losses (MW) | 13.29 | 9.29 | 16.89 | 13.16 |

3) FDI minimally changes the true electrical state but significantly alters the perceived one, underscoring the risk of measurement-centric attacks.

## VI. DISCUSSION

The results in Section V show that even brief, localized cyberattacks can meaningfully alter both local and system-wide behavior in the IEEE 14-bus network. Bus 5 plays a particularly influential role due to its electrical proximity to GEN 6, which gives it strong leverage over nearby voltage profiles. When DoD falsifies the load at this bus, the distorted power balance triggers reactive adjustments from adjacent generators, producing the clustered overvoltages observed at buses 5–9. This demonstrates that manipulation at a single, well-connected node can propagate through tightly coupled branches and affect broader system stability.

DoS attacks exhibit a different pattern: by freezing $P_d$ and removing natural load variation, they produce flatter voltage traces that give a false sense of stability. Although the physical system continues to evolve, operators observe stale telemetry, which can delay the detection of genuine disturbances or load changes. FDI attacks, in contrast, primarily introduce informational risk. By altering reported $V/P/Q$ values without significantly changing the physical state, they create a divergence between "true" and "measured" conditions. This enables both masking of real violations and the creation of false alarms, with only modest changes in RMS deviation.

Reactive-limit behavior further highlights system stress under attack. DoD increases PV→PQ transitions as generators hit their reactive power limits more frequently, reducing voltage regulation capability and shrinking stability margins. Under coordinated attacks, this loss of reactive flexibility could accelerate voltage instability even when individual telemetry signals appear nominal. Overall, the results demonstrate that cyberattacks reshape both operator perception and underlying physical response, with localized perturbations capable of triggering broader system effects. This work focuses on an offline simulation environment and therefore does not explicitly model communication delays, packet loss, or network congestion. While this abstraction enables controlled and repeatable analysis of cyber–physical interactions, it limits direct representation of real-time grid operations; these aspects are left for future extensions of the framework.

## VII. CONCLUSION

This paper introduced a MATLAB-based, time-stepped framework for assessing communication-layer cyberattacks on the IEEE 14-bus system. By combining realistic load variation, attack injection, and full AC power flow with reactive-limit enforcement, the approach captures cyber–physical effects that static analyses overlook. The results reveal clear distinctions between attack types: DoS suppresses natural variability and reduces situational awareness, DoD produces the largest physical disturbances and reactive-power stress, and FDI primarily distorts perceived voltages while masking or fabricating violations. Overall, the study demonstrates how brief telemetry manipulation can propagate through power-flow dynamics and mislead operator decision-making, highlighting the need for resilient monitoring and detection under compromised conditions. The proposed framework serves as a reproducible basis for systematic cyber–physical experimentation. Future work will extend the framework to incorporate realistic communication delays, packet loss, and asynchronous measurement updates to better reflect operational environments. Integration with state estimation and automatic voltage control will enable evaluation of closed-loop cyber–physical interactions under attack. In addition, the generated telemetry streams provide a foundation for testing anomaly detection and intrusion detection techniques. Finally, scaling the approach to larger transmission systems will support assessment of how attack impacts evolve with increased network size and complexity.